\documentclass[aps,amsfonts,amsmath,prd,preprint,nofootinbib]{revtex4}

\usepackage[utf8]{inputenc}
\usepackage{amsfonts}
\usepackage{physics}
\usepackage{mathtools}
\usepackage{mathrsfs}
\usepackage{amssymb}
\usepackage{appendix}
\usepackage{graphicx}
\usepackage{amsmath} 
\usepackage{color}
\usepackage{feynmf}
\usepackage{simpler-wick} 
\usepackage[legalpaper, margin=1in]{geometry}
\usepackage{soul}

\begin{document}

\title{What determines the rest frame of bubble nucleation?}

\author{Yilin Chen and Alexander Vilenkin}
\address{Institute of Cosmology, Department of Physics and Astronomy, \\
Tufts University, Medford, Massachusetts 02155, USA}

\begin{abstract}
We revisit the question addressed in recent papers by Garriga et al: What determines the rest frame of pair nucleation in a constant electric field?  The conclusion reached in these papers is that pairs are observed to nucleate at rest in the rest frame of the detector which is used to detect the pairs.  A similar conclusion should apply to bubble nucleation in a false vacuum.  This conclusion however is subject to doubt due to the unphysical nature of the model of a constant eternal electric field that was used by Garriga et al.  The number density of pairs in such a field would be infinite at any finite time.  Here we address the same question in a more realistic model where the electric field is turned on at a finite time $t_0$ in the past.  The process of turning on the field breaks the Lorentz invariance of the model and could in principle influence the frame of pair nucleation.  We find however that the conclusion of Garriga et al still holds in the limit $t_0 \to -\infty$.  This shows that the setup process of the electric field does not have a lasting effect on the observed rest frame of pair nucleation.  {On the other hand, the electric current and charge density due to the pairs are determined by the way in which the electric field was turned on.}

\end{abstract}

\maketitle

\section{Introduction}

Coleman's description of vacuum decay through bubble nucleation \cite{coleman} is manifestly Lorentz invariant.  The tunneling instanton is $O(4)$ invariant and its analytic continuation is invariant under Lorentz boosts.  There is however some tension between this description and the semiclassical picture of bubble nucleation. The analytically continued instanton describes a bubble contracting relativistically from infinite size, then bouncing and re-expanding back to infinity. On the other hand, physically we expect a bubble to nucleate at rest and expand, so the contracting part of the worldsheet should be cut off \cite{Voloshin}.  But then the cutoff breaks the Lorentz invariance of the bubble worldsheet, and we have to face the question: What determines the rest frame of bubble nucleation?

One can anticipate two possible scenarios: (A) It could be that the frame of nucleation simply coincides with the rest frame of the detector used to observe the bubble. In other words, each observer will see bubbles forming at rest in her own rest frame.  (B) Another possibility is that the frame of nucleation is influenced by how the decaying false vacuum was set up. If the false vacuum has zero energy, its spacetime is flat, and the space will be filled by nucleating bubbles in a finite amount of time. This implies that the false vacuum could not have existed forever; it must have been created in some manner in the past.  

This issue was first addressed by Garriga \textit{et al} in Ref. \cite{Garriga1} using the close analogy between bubble nucleation and pair production in an electric field.  They considered a charged scalar field $\phi$ in a constant electric field $E$ in $(1+1)$ dimensions.  The field was assumed to be in the in-vacuum state which was prepared in the infinite past.  
It was shown in \cite{Garriga1} that this quantum state is Lorentz invariant.
In addition to $\phi$, Ref.~\cite{Garriga1} introduced another charged field $\psi$ and a real field $\chi$ which play the role of a detector.  The interaction between the fields was chosen of the form\footnote{This model was earlier studied by Massar and Parentani \cite{Massar} and by Gabriel {\it et al.} \cite{Gabriel} to investigate the Unruh effect for an accelerated detector.}
\begin{align}
{\cal H}_{int}=g(\phi\psi^\dagger \chi +{\rm h.c.}), 
\label{Lint}
\end{align}
where $g$ is a coupling constant.  If a $\psi$-particle is present in the initial state, it can annihilate a $\phi$-antiparticle via
$\psi{\bar\phi}\to\chi$.  One can then study the momentum distribution of $\chi$-particles in the final state to deduce the momentum distribution of the created pairs. It was found in Ref.~\cite{Garriga1} that particles (antiparticles) of the pairs are predominantly observed moving in the direction of (opposite to) the electric field. 
This was interpreted to indicate that option (A) above is correct: the pairs are mostly observed to nucleate in the rest frame of the detector.  This conclusion was later reinforced in Refs.~\cite{Garriga2,YCAV}.


A potential problem with these results is that, as noted in Ref.~\cite{Garriga1}, the in-vacuum state in a constant electric field has some unphysical properties.  The singularity structure of the two-point function in this state does not have the Hadamard form, and as a result the expectation values of physical observables cannot be regulated in a Lorentz invariant way.  An important special case is that of the electric current. One expects that the created pairs moving in the electric field will develop a nonzero current, which will break the Lorentz invariance. And indeed one can show that all physical (Hadamard) states of charged particles in a constant electric field are not Lorentz invariant \cite{Garriga1}.  

Physically, this issue is related to the fact that a metastable vacuum could not have existed for an infinite time.
With pairs created at a constant rate starting from infinite past, the density of $\phi$-particles at any finite time would be infinite.
In a more physical approach the electric field would have to be time-dependent with $E(t\to -\infty)\to 0$, so the initial state of the field $\phi$ could be chosen as the standard vacuum state.  The two-point function is then Hadamard at $t\to -\infty$ and is guaranteed to remain Hadamard at later times.  The time dependence of the electric field introduces a preferred frame and explicitly breaks the Lorentz invariance.  The question is to what extent this Lorentz violation influences the conclusions of \cite{Garriga1, Garriga2}. This question is at the focus of the present paper.  

We are going to follow the method of Refs.~\cite{Garriga1,Garriga2}, except that a constant electric field $E$ is turned on at some early time $t_0$ and is set to zero at $t<t_0$.  
In the next section we specify the quantum states of the pair-producing field $\phi$ and of the detector fields $\psi$ and $\chi$.   
Following Ref.~\cite{Garriga1}, in order to pinpoint the time of detection (and the detector rest frame at that time), we switch the detector on for a short time interval $T$ around $t=0$.  This is implemented by introducing a time-dependent coupling $g(t)=g e^{-t^2/T^2}$ in Eq.(\ref{Lint}).  
In Section 3 we analyze the pair detection amplitude and compare it with that for an eternal electric field. 
In Section 4 we perform a similar analysis using an alternative, neutral detector model introduced in Ref.~\cite{Garriga2}.  We shall see that the latter model offers significant advantages; in particular, the time-dependent coupling $g(t)$ is no longer needed.  The results we obtain with the two detector models are fully consistent with one another and with the earlier results of Refs.~\cite{Garriga1,Garriga2}. Our conclusions are summarized and discussed in Section 5.



\section{Physical setup}

We shall consider an electric field
\begin{align}
    E(t)=
    \begin{cases}
     &0,\ t<t_0, \\
     &E,\ t\geq t_0.
    \end{cases}
\end{align}
Pair creation in a similar setting has been discussed by Adorno, Gavrilov and Gitman \cite{Gitman2}, the main  difference being that they studied an electric field which is present for a finite interval of time and is turned off at some  $t_1>t_0$.  Here we do not need to introduce a turn-off time.

The hypersurface $t=t_0$ defines an inertial reference frame $\Sigma_0$, and our goal here will be to determine to what extent pair nucleation is influenced by the existence of this frame.  {We note that such influence is not {\it a priori} excluded, even in the limit of $t_0\to -\infty$.  One example is the "persistence of memory" effect on bubble collisions \cite{persis}.}

Particle-antiparticle pairs are described by a complex field operator $\phi(x,t)$,
\begin{align}\label{phiexp}
  \phi(x,t)\!&=\!\int\frac{dk}{\sqrt{2\pi}}\left(a_k \phi_k(t)+b^{\dagger}_{-k} \phi^{*}_k(t)\right)e^{ikx},  
\end{align}  
where $a_k$ and $b_k$ are respectively the particle and antiparticle annihilation operators in the in-vacuum state $\ket{0,in}$,
\begin{align}
a_k\ket{0,in}=b_k\ket{0,in}=0.
\end{align}
We also introduce two additional fields, $\psi$ and $\chi$, which will play the role of a detector.  $\psi$ is a complex field, 
\begin{align}
    \psi(x,t) \!&=\!\int\frac{dq}{\sqrt{2\pi}}\left(d_{q}\psi_{q}(t)
    +f^{\dagger}_{-q}\psi^{*}_{q}(t)	\right)e^{iqx},
\end{align}
and $\chi$ is real,
\begin{align}
    \chi(x,t)\!&=\!\int\frac{dp}{\sqrt{2\pi}}\left(c_{p} \chi_{p}(t)+c^{\dagger}_{-p}\chi^*_{p}(t)\right)e^{ipx}.
\end{align}
In what follows, we shall always use the notation $k$, $q$ and $p$ to denote the momenta of $\phi$, $\psi$ and $\chi$ particles, respectively.  We shall also use $m_\phi$, $m_\psi$ and $m_\chi$ for the masses of $\phi$, $\psi$ and $\chi$-particles respectively.

At $t<t_0$, the mode functions $\phi_k(t)$ corresponding to the in-vacuum state are positive-frequency plane-wave solutions of the Klein-Gordon equation, while at $t>t_0$ they are linear combinations of solutions in a constant external electric field.\footnote{ Here we use the gauge $A_0=0$.}  The particular combination can be found by matching $\phi_k(t)$ and their time derivatives at $t=t_0$ \cite{Gitman2}:
\begin{align}\label{phi}
    \phi_k(t)=
    \begin{cases}
     C(\Omega_k)\exp[-i\Omega_k (t-t_0)],\ \ &t<t_0,\\
     C(\Omega_k)\left\{a_1(\Omega_k) D_{\nu^*}[-(1-i)z]+a_2(\Omega_k)D_{\nu}[-(1+i)z]\right\},\ \ &t\geq t_0,
    \end{cases}
\end{align}
where
\begin{align}
    \Omega_k=\sqrt{m^2_{\phi}+(k+eEt_0)^2},\ \ z=\sqrt{eE}\left(t+\frac{k}{eE}\right),\ \ \nu=-\frac{1+i\lambda}{2},\ \ \lambda=\frac{m_{\phi}^2}{eE},
\end{align}
$C(\Omega_k)=1/\sqrt{2\Omega_k}$ and $D_\nu$ are parabolic cylinder functions.
The coefficients $a_1$ and $a_2$ satisfy 
\begin{align}\label{a1a2}
    C^2(\abs{a_1}^2-\abs{a_2}^2) = \frac{e^{-\frac{\pi}{4}\lambda}}{\sqrt{2eE}}
\end{align}
and are given by
\begin{align}\label{coeff}
    a_j = \frac{(-1)^j}{\sqrt{2}}e^{\frac{i\pi}{2}\left(\frac{1}{2}+\nu^*\right)}\sqrt{z^2_0+\lambda}f_j^{(+)}(z_0), \ \ \ \ j=1,2,
\end{align}
where\footnote{Note the sign difference with \cite{Gitman2} in the terms containing derivatives. This is because the mode functions in \cite{Gitman2} are negative frequency modes, while we are using positive frequency modes.}
\begin{align}
    f_1^{(+)}(z)=\left(1-\frac{i}{\sqrt{z^2+\lambda}}\frac{d}{dz}\right)D_{\nu}[-(1+i)z],\nonumber\\
    f_2^{(+)}(z)=\left(1-\frac{i}{\sqrt{z^2+\lambda}}\frac{d}{dz}\right)D_{\nu^*}[-(1-i)z]
\end{align}
and $z_0=z(t_0)$.
The relation (\ref{a1a2}) follows from the definitions (\ref{coeff}) and the Jacobian
\begin{align}
    D_{\nu}(z)\frac{d}{dz}D_{-1-\nu}(iz)-D_{-1-\nu}(iz)\frac{d}{dz}D_{\nu}(z) = -ie^{-i\frac{\pi}{2}\nu}.
\end{align}
In the case of a time-independent (eternal) electric field considered in \cite{Garriga1}, the in-vacuum mode functions are obtained by setting $a_2=0$ in Eq.(\ref{phi}).

The mode functions $\psi_q$ for $\psi$-particles are given by similar expressions with $k$ replaced by $q$ and $m_\phi$ by $m_\psi$. We shall denote the coefficients $a_j$ for $\phi$ and $\psi$ particles by $a_{j\phi}$ and $a_{j\psi}$ respectively. For neutral $\chi$-particles the mode functions are simply plane waves,
 \begin{align}\label{chi}
    \chi_p(t)= \frac{1}{\sqrt{2\omega_p}}e^{-i\omega_p t},\ \ \ \ \omega_p=\sqrt{p^2+m^2_{\chi}}.
\end{align}

The detection process is a scattering $\psi\phi^*\to\chi$ in which a $\psi$-particle detector collides with the $\phi$-antiparticle of the pair, turning into a neutral $\chi$-particle.  Our goal is to study the correlation between the observed frame of pair nucleation and the rest frame of the $\psi$-detector at the time of collision.  Note however that the $\psi$-particle is accelerated by the electric field, so its rest frame changes with time.  
In order to pinpoint the time of collision, we make the coupling $g$ in Eq.(\ref{Lint}) time-dependent:
\begin{align}
g(t)=g e^{-t^2/T^2}.
\end{align}
This ensures that the detection occurs in the interval $\Delta t\sim T$ around $t=0$. 
We shall assume that 
\begin{align}\label{ine1}
t_0<0,~~~ |t_0|\gg T,
\end{align}
so that the measurement is well separated from the setup of the electric field.



It is shown in \cite{Garriga1} that the momentum distribution of $\chi$-particles after the measurement can be expressed as
 \begin{align}
 \frac{dN_\chi}{dp}=\frac{1}{2\pi}\abs{{\cal A}_\chi(p;q)}^2,
 \end{align}
 where the amplitude ${\cal A}_\chi$ is given by
\begin{align}\label{Apq}
    \mathcal{A}_{\chi}(p;q) = \int dt\ g(t)\phi^*_{q-p}(t)\psi_q(t)\chi^*_p(t).
\end{align}
In the next section we will analyze this integral and compare the result with that for an eternal electric field.

\section{Amplitude for the process $\psi{\bar\phi}\to\chi$}

We will be interested in the amplitude (\ref{Apq}) in the limit $t_0\to -\infty$. Substituting the mode functions \eqref{chi}, \eqref{phi} and similar mode function for $\psi$-particles in Eq.(\ref{Apq}), we express the amplitude as a sum of five integrals:
\begin{align}\label{amp}
    \mathcal{A}_{\chi}(p;q)=I_0+I_{11}+I_{12}+I_{21}+I_{22},
\end{align}
where
\begin{align}
    I_0&=\frac{g}{2\sqrt{2\omega_p\xi_q\Omega_{q-p}}}\int_{-\infty}^{t_0}dt\ e^{i(\Omega_{q-p}-\xi_q+\omega_p)t-t^2/T^2},\\
    I_{11}&=\frac{g a^*_{1\phi}a_{1\psi}}{2\sqrt{2\omega_p\xi_q\Omega_{q-p}}}\int_{t_0}^{\infty}dt\ e^{i\omega_p t-t^2/T^2}D_{\nu_{\phi}}[-(1+i)z_{\phi}]D_{\nu^*_{\psi}}[-(1-i)z_{\psi}],\\
    I_{12}&=\frac{g a^*_{1\phi}a_{2\psi}}{2\sqrt{2\omega_p\xi_q\Omega_{q-p}}}\int_{t_0}^{\infty}dt\ e^{i\omega_p t-t^2/T^2}D_{\nu_{\phi}}[-(1+i)z_{\phi}]D_{\nu_{\psi}}[-(1+i)z_{\psi}],\\
    I_{21}&=\frac{g a^*_{2\phi}a_{1\psi}}{2\sqrt{2\omega_p\xi_q\Omega_{q-p}}}\int_{t_0}^{\infty}dt\ e^{i\omega_p t-t^2/T^2}D_{\nu^*_{\phi}}[-(1-i)z_{\phi}]D_{\nu^*_{\psi}}[-(1-i)z_{\psi}],\\
    I_{22}&=\frac{g a^*_{2\phi}a_{2\psi}}{2\sqrt{2\omega_p\xi_q\Omega_{q-p}}}\int_{t_0}^{\infty}dt\ e^{i\omega_p t-t^2/T^2}D_{\nu^*_{\phi}}[-(1-i)z_{\phi}]D_{\nu_{\psi}}[-(1+i)z_{\psi}].
\end{align}
Here, $\xi_q = \sqrt{(q+eEt_0)^2+m^2_{\psi}}$, $z_{\phi}=\sqrt{eE}\left(t+\frac{q-p}{eE}\right)$ and $z_{\psi}=\sqrt{eE}\left(t+\frac{q}{eE}\right)$. 

The first integral $I_0$ is rather simple
\begin{align}
    I_0=\frac{\sqrt{\pi}gT}{4\sqrt{2\omega_p\xi_q\Omega_{q-p}}}e^{-\frac{1}{4}(\Omega_{q-p}-\omega_q+\omega_p)^2T^2}\left(1- \text{erf}\left(\frac{i(\Omega_{q-p}-\omega_q+\omega_p)T}{2}-\frac{t_0}{T}\right)\right),
\end{align}
where $\text{erf}$ is the error function.  Due to the assumption  \eqref{ine1}, we can use the asymptotic form
\begin{align}\label{erf}
    \text{erf}(z) \approx 1-\frac{e^{-z^2}}{\sqrt{\pi}z} \ \ \ \ \left(\abs{z}\gg 1,~~\abs{\arg{z}}<\frac{3\pi}{4}\right)
\end{align}
This shows that $I_0$ approaches zero exponentially fast as $-t_0/T\to\infty$: 
\begin{align}\label{I0exp}
I_0\propto \exp\left(-\frac{t_0^2}{T^2}\right). 
\end{align}

Now, it is shown in Appendix A that the lower limit of integration in the remaining integrals $I_{jk}$ can be extended from $t_0$ to $-\infty$ with an error which is also $O\left[\exp\left(-\frac{t_0^2}{T^2}\right)\right]$.  The dependence on $t_0$ in $I_{jk}$ is then limited to the coefficients $a_j(\phi)$, $a_j(\psi)$.  Furthermore, in Appendix B we show that 
\begin{align}
\frac{a_2}{a_1}\sim \frac{1}{eE t_0^{2}} .
\end{align}
for both $\phi$ and $\psi$ modes.  We can therefore neglect $I_{12}$, $I_{21}$ and $I_{22}$ compared to $I_{11}$ in the limit of $t_0\to -\infty$.  Finally, it follows from Eq.(\ref{a1a2}) that in this limit the $a_1$ coefficients have the same values as for an eternal electric field (when $a_2=0$).  

This shows that the amplitude for the $\phi^*$-particle detection in a field turned on at a finite $t_0$ approaches the amplitude for a constant eternal field in the limit $t_0\to -\infty$. The conclusion is that, as far as these detection processes are concerned, the in-vacuum state gradually "forgets" how it was created and measurements of pair nucleation at late times yield a nucleation frame close to the rest frame of the detector.

An attentive reader might have noticed that there is still a potential problem with this analysis.
We used a time-dependent coupling $g(t)$, where $t$ is the time coordinate in the frame $\Sigma_0$ of initial conditions. This time dependence also breaks Lorentz invariance and could in principle influence the frame of nucleation.  One could argue that our result, that the detection amplitude is independent of $\Sigma_0$, indicates that it is also independent of how the electric field was set up.  There is however still a caveat that the two Lorentz violations could somehow compensate one another.  This seems rather unlikely, but it would be better to have a cleaner analysis where this issue would not arise.  In the next Section we will show that such analysis can be performed using the neutral detector model of Ref.~\cite{Garriga2}.

\section{Neutral detector}

The neutral detector model of Ref.~\cite{Garriga2} is based on the 4-point interaction 
\begin{align}
    {\cal H}_{int} = g(\chi_1\chi^*_2\phi\psi^*+\chi^*_1\chi_2\phi^*\psi),
\end{align}
where $\chi_1$ and $\chi_2$ are electrically neutral complex scalar fields.  The detection process is the scattering $\chi_1 \phi \to \chi_2 \psi$.  The scalar detector particle $\chi_1$ collides with the $\phi$ particle of a pair producing a different kind of scalar particle $\chi_2$ and a charged $\psi$-particle.  By measuring the momenta of the incoming $\chi_1$ and outgoing $\chi_2$ particles one can determine the momentum of the $\phi$ particle at the time of collision \cite{Garriga2}.  The benefit of this setup is clear: we will not need to turn the interaction on and off since the momenta of the detector and the product are not affected by the electric field.

It is shown in Ref.~\cite{Garriga2} that the momentum distribution of $\chi_2$-particles after the measurement is 
\begin{align}
    \frac{dN_2}{dqdp} = \frac{1}{(2\pi)^2}\abs{\mathcal{A}_2(k, q, {p};{p}')}^2,
\end{align}
where the amplitude is given by
\begin{align}
    \mathcal{A}_2(k, q, {p};{p}')=g\int_{-\infty}^{\infty} dt\ \phi^*_k(t)\psi^*_q(t)\chi_{1,{p}'}(t)\chi^*_{2,{p}}(t)
\label{A2}
\end{align}
with $k=q+p$.  Here, $p'$ and $p$ are respectively the momenta of $\chi_1$ and $\chi_2$, while $k$ and $q$ are respectively the canonical momenta of $\phi$ and $\psi$.

To compare the amplitude (\ref{A2}) to that for the same process in an eternal electric field, we will closely follow our analysis of the amplitude (\ref{Apq}) for a charged detector.  As in Eq.(\ref{amp}), the amplitude can be represented as a sum of five integrals.  The main difference is that the integrals now do not include the Gaussian suppression factors $\exp(-t^2/T^2)$.  Integration can again be extended to the infinite range (the details are given in Appendix C) and terms including the coefficients $a_{2,\phi}$, $a_{2,\psi}$ can be neglected with an error $\propto t_0^{-2}$.  Then the amplitude (\ref{A2}) reduces to that in an eternal electric field.

The latter amplitude was calculated by Garriga {\it et al} \cite{Garriga2}, who found that it is independent of $q$ and that the distribution of $\chi_2$ particles can be expressed as
\begin{align}
    \frac{dN_2}{dtdp} = \frac{eE}{(2\pi)^2}\abs{\mathcal{A}_2({p};{p}')}^2.
\end{align}
Furthermore, in the frame of the $\chi_1$ detector $(p'=0)$ they found that $\phi$ particles are moving predominantly in the direction of $E$.  Thus the pairs are observed to nucleate in the detector frame.

\section{Summary and discussion}

We have revisited the question that was addressed
in a number of recent papers \cite{Garriga1,Garriga2,YCAV}: What determines the rest frame of pair nucleation in a constant electric field?  The conclusion reached in Refs. \cite{Garriga1,Garriga2,YCAV} is
that pairs are observed to nucleate at rest in the rest frame of the detector which is used to detect the pairs.  The physics of pair production is Lorentz invariant, but the invariance is broken by the interaction of pairs with the detector.

As we discussed in the Introduction, this conclusion is subject to doubt because the quantum state of the pairs that was used in Refs.\cite{Garriga1,Garriga2,YCAV} assumed that a constant electric field had existed forever and therefore had some unphysical properties.  In particular, the number density of pairs in such a field would be infinite at any finite time.  In a physically realistic setting, the electric field would be turned on at a finite time in the past.  This, however, would also break Lorentz invariance and could in principle influence the frame of pair nucleation.
Our goal in this paper was to resolve this remaining ambiguity.   We used a $(1+1)$-dimensional model of pair production in electric field very similar to that in Ref.\cite{Gitman2}.  We also used the same model of particle detector. The only difference is that in our present model the electric field is turned on at a finite time $t_0$ in the distant past.  We found that the amplitude for particle detection approaches that in a constant eternal field in the limit $t_0\to -\infty$. This shows that the setup process of the electric field does not have a lasting effect on the observed rest frame of pair nucleation.  

We thus conclude that at late times, after the initial disturbance caused by turning on the electric field has settled down, the pairs are observed to nucleate with their rest frame close to the rest frame of the detector.
Due to the close analogy between pair production and bubble nucleation in false vacuum, we expect a similar conclusion to apply in the latter case as well: all inertial observers will observe bubbles to nucleate in their own rest frames.   

We emphasize however that the way in which the false vacuum is set up does have an effect on some physical observables.  In the case of our pair nucleation model this can be illustrated by the $\phi$-particle electric current $J^\mu(t)$.  The expectation value of this current in the quantum state (\ref{phi}) was calculated in Ref.\cite{Gitman2}.  In the late time limit\footnote{Here, the back-reaction of the pairs on the electric field is neglected, as it was done in \cite{Garriga1,Garriga2} due to the weak field limit $m^2\gg eE$. For weak fields, the back-creation takes place on a long
timescale, which should be at least as long as $t\sim \dot n^{-1/d}$, where $d$ is the spacetime dimension. A thorough computation of Fermion can be found in Ref.~\cite{fermion}.} it is given by
\begin{align}
J^0=0,~~~J^1\approx C(t-t_0),
\end{align}
where $C=2e{\dot n}_\phi$ and
\begin{align}
{\dot n}_\phi=\frac{e^2 E}{\pi} \exp\left(-\frac{\pi m_\phi^2}{eE}\right)
\end{align}
is the rate of $\phi$-pair production.  This result has a clear physical meaning.  Particles and antiparticles are produced at the same rate, hence the charge density is $J^0=0$.  The pairs rapidly approach relativistic speeds, and thus the current density grows at the rate $dJ^1/dt \sim 2e{\dot n}_\phi$.  This picture, however, applies only in a specific reference frame, where the electric field is turned on at the same time $t_0$ everywhere in space.  In a different reference frame, moving at velocity $v$ with respect to the original one, the current is obtained by a Lorentz transformation:
\begin{align}
{J^1}' = \gamma^2 C (t' + vx'),~~~{J^0}' = - v {J^1}',
\label{J'}
\end{align}
where $\gamma=(1-v^2)^{-1/2}$ is the Lorentz factor.  The frame in which the electric field was set up can be determined by measuring, for example, the charge density $J^0$: it is the frame where $J^0=0$.  This memory of the initial state persists no matter how much time elapsed after the field was turned on.

The nature of this effect is similar to that of the "persistence of memory" effect on bubble nucleation \cite{persis}.  An observer in a false vacuum will generally see nucleating bubbles arriving at different rates from different directions.  Once again, this anisotropy will persist at arbitrarily late times.  The distribution of the arrival directions will be isotropic only for an observer whose world line is orthogonal to the hypersurface where the false vacuum was set up.  The corresponding effect in our present context is the following.  Particles and antiparticles move in opposite directions at nearly the speed of light and generally have different densities.  This means that they generally have different fluxes.  The fluxes will be the same only in the frame where the electric field was set up.   

\acknowledgements

A.V. is grateful to Jaume Garriga for useful discussions.  This work was supported in part by the National Science Foundation under grant No. 2110466.

\appendix

\section{Extending integration to $t\to -\infty$}

Here we show that integration over $t$ in the integrals $I_{11}$, $I_{12}$, $I_{21}$, $I_{22}$ can be extended to $t\to -\infty$ with an exponentially small error.  

Taking $I_{11}$ as an example, we can express it as
\begin{align}
    I_{11}&=& \frac{g a^*_{1\phi}a_{1\psi}}{2\sqrt{2\omega_p\xi_q\Omega_{q-p}}}\left\{\int_{-\infty}^{\infty}-\int_{-\infty}^{t_0}\right\}dt\ e^{i\omega_p t-t^2/T^2}D_{\nu_{\phi}}[-(1+i)z_{\phi}]D_{\nu^*_{\psi}}[-(1-i)z_{\psi}] \equiv I - I',
\end{align}
where $I$ is the extended integral over the infinite range and $I'$ is the subtracted remnant.  We need to check that the remnant $I'$ is negligible.

 Since we are interested in the limit $t_0\to -\infty$ and the integration in $I'$ is restricted to the interval $(-\infty, t_0]$, we can use the asymptotic form of the parabolic cylinder functions:
 \begin{align}
 D_\nu(z)\sim z^\nu e^{-\frac{z^2}{4}}.
 \label{asympD}
 \end{align} 
 Then we have
 \begin{align}
    I'\propto \int_{-\infty}^{t_0}dt\ e^{i\omega_p t-t^2/T^2}\left(\sqrt{2}\abs{z_{\psi}}\right)^{\nu_{\psi}^*} e^{\frac{i}{2}z_{\psi}^2}\left(\sqrt{2}\abs{z_{\phi}}\right)^{\nu_{\phi}} e^{-\frac{i}{2}z_{\phi}^2}.
\end{align}
This can be further simplified using $z_\phi, z_\psi \approx \sqrt{eE}t$, and we find 
\begin{align}
 \abs{I'} \propto \abs{\int_{-\infty}^{t_0}dt\ e^{-t^2/T^2}\left(-\sqrt{2eE}t\right)^{\nu_{\psi}^*+\nu_{\phi}}}
 < \left(\sqrt{2eE}|t_0|\right)^{Re(\nu_{\psi}^*+\nu_{\phi})}\int_{-\infty}^{t_0}dt\ e^{-t^2/T^2}\nonumber\\
    =  \left(\sqrt{2eE}|t_0|\right)^{-1}\frac{\sqrt{\pi}T}{2}e^{-\frac{\omega_p^2T^2}{4}}\left[1-\text{erf}\left(-\frac{t_0}{T}\right)\right],
\end{align}
where we have used the fact that $Re(\nu_\psi^*+\nu_\phi)=-1$.  Now, using Eq.\eqref{erf}, we see that $I'$ approaches zero exponentially:
\begin{align}
    \abs{I'} \propto \exp\left(-\frac{t^2_0}{T^2}\right).
\end{align}
The same argument can be applied to the remaining integrals $I_{12}$, $I_{21}$, $I_{22}$.

\section{Relative magnitude of $a_1$ and $a_2$}

From Eq.\eqref{coeff} we know that $a_j  \propto  f_j^{(+)} (z_0)$.  In the limit $t_0\to -\infty$ we can use
the asymptotic form (\ref{asympD}) of the parabolic cylinder functions to evaluate $f_j^{(+)}$.
Starting with $f_1^{(+)}$, we can write
\begin{align}
    f_1^{(+)}(z_0)&=\left(1-\frac{i}{\sqrt{z^2+\lambda}}\frac{d}{dz_0}\right)D_{\nu}[-(1+i)z_0]\nonumber\\
    &\approx \left(1-\frac{z_0}{\sqrt{z_0^2+\lambda}}\right)e^{-i\frac{z^2_0}{2}}[-(1-i)z_0]^{\nu}.
\end{align}
Then, with $z_0\to -\infty$ we have
\begin{align}
   |f_1^{(+)}(z_0\to -\infty)| \sim |z_0|^{-1/2}
\end{align}

For $f_2^{(+)}(z_0)$ we need to keep terms up to second order:
\begin{align}
    f_2^{(+)}(z_0)&\approx \left(1+\frac{z_0}{\sqrt{z_0^2+\lambda}}-\frac{(1+i)\nu^*}{z_0\sqrt{z_0^2+\lambda}}\right)e^{i\frac{z^2_0}{2}}[-(1+i)z_0]^{\nu^*}\nonumber\\
    &\approx \frac{(1+i)\nu^*}{z^2_0}e^{i\frac{z^2_0}{2}}[-(1+i)z_0]^{\nu^*}
    \end{align}
We thus conclude that
\begin{align}
   |f_2^{(+)}(z_0\to -\infty)| \sim |z_0|^{-5/2}
\end{align}
and
\begin{align}
\abs{\frac{a_2}{a_1}}\sim |z_0|^{-2}\sim\frac{1}{eE t_0^{2}} .
\end{align}

\section{Amplitude for a neutral detector}

Following the analysis for a charged detector in Section 3, we express the amplitude (\ref{A2}) as a sum of five terms:
\begin{align}
    \mathcal{A}_2(k, q, {p};{p}')=I_0+I_{11}+I_{12}+I_{21}+I_{22}.
\end{align}
Here
\begin{align}
    I_0 &= \frac{g}{4(\omega_{\Tilde{p}}\omega_{\Tilde{p}'}\xi_q\Omega_k)^{1/2}}\int_{-\infty}^{t_0}dt\ e^{i(\xi_q + \Omega_k + \omega_{\Tilde{p}} - \omega_{\Tilde{p}'})t-\epsilon t}, \label{C2}  \\ 
    I_{11}&=\frac{g a^*_{1\phi}a^*_{1\psi}}{4(\omega_{\Tilde{p}}\omega_{\Tilde{p}'}\xi_q\Omega_k)^{1/2}}\int_{t_0}^{\infty}dt\ e^{i(\omega_{\Tilde{p}} - \omega_{\Tilde{p}'})t}D_{\nu_{\phi}}[-(1+i)z_{\phi}]D_{\nu_{\psi}}[-(1+i)z_{\psi}],\\
    I_{12}&=\frac{g a^*_{1\phi}a^*_{2\psi}}{4(\omega_{\Tilde{p}}\omega_{\Tilde{p}'}\xi_q\Omega_k)^{1/2}}\int_{t_0}^{\infty}dt\ e^{i(\omega_{\Tilde{p}} - \omega_{\Tilde{p}'})t}D_{\nu_{\phi}}[-(1+i)z_{\phi}]D_{\nu^*_{\psi}}[-(1-i)z_{\psi}],\\
    I_{21}&=\frac{g a^*_{2\phi}a^*_{1\psi}}{4(\omega_{\Tilde{p}}\omega_{\Tilde{p}'}\xi_q\Omega_k)^{1/2}}\int_{t_0}^{\infty}dt\ e^{i(\omega_{\Tilde{p}} - \omega_{\Tilde{p}'})t}D_{\nu^*_{\phi}}[-(1-i)z_{\phi}]D_{\nu_{\psi}}[-(1+i)z_{\psi}],\\
    I_{22}&=\frac{g a^*_{2\phi}a^*_{2\psi}}{4(\omega_{\Tilde{p}}\omega_{\Tilde{p}'}\xi_q\Omega_k)^{1/2}}\int_{t_0}^{\infty}dt\ e^{i(\omega_{\Tilde{p}} - \omega_{\Tilde{p}'})t}D_{\nu^*_{\phi}}[-(1-i)z_{\phi}]D_{\nu^*_{\psi}}[-(1-i)z_{\psi}].
\end{align}
and we have inserted a convergence factor $\exp(-\epsilon t)$ with $\epsilon\to 0+$ in Eq.(\ref{C2}). 

The integration in $I_0$ is straightforward; we have
\begin{align}\label{4I0}
    I_0 &= \frac{-ig}{4(\omega_{\Tilde{p}}\omega_{\Tilde{p}'}\xi_q\Omega_k)^{1/2}}\frac{e^{i(\xi_q + \Omega_k + \omega_{\Tilde{p}} - \omega_{\Tilde{p}'})t_0}}{\xi_q + \Omega_k + \omega_{\Tilde{p}} - \omega_{\Tilde{p}'}}
\end{align}    
and
\begin{align}    
    \abs{I_0} \propto (\sqrt{eE}\abs{t_0})^{-2}
\end{align}
with a coefficient independent of $t_0$.

As before, we can neglect $I_{12}$, $I_{21}$ and $I_{22}$ compared to $I_{11}$ in the limit of $t_0\to -\infty$. Further, the integration in $I_{11}$ can be extended to $- \infty$ with a small error. 
To verify this we write
\begin{align}
    I_{11}=&\frac{g a^*_{1\phi}a^*_{1\psi}}{4(\omega_{\Tilde{p}}\omega_{\Tilde{p}'}\xi_q\Omega_k)^{1/2}}\left\{\int_{-\infty}^{\infty}-\int_{-\infty}^{t_0}\right\}dt\ e^{i(\omega_{\Tilde{p}} - \omega_{\Tilde{p}'})t}D_{\nu_{\phi}}[-(1+i)z_{\phi}]D_{\nu_{\psi}}[-(1+i)z_{\psi}]\nonumber\\
    \equiv& I-I',
\end{align}
where $I$ is the extended integral over the infinite range and $I'$ is the subtracted remnant.  We need to check that the remnant $I'$ is negligible.  As in Appendix A, we can use $z_\phi, z_\psi \approx \sqrt{eE}t$ and the 
asymptotic form \eqref{asympD} of the parabolic cylinder functions:
\begin{align}
    I'\propto \int_{-\infty}^{t_0}dt\ e^{i(\omega_{\Tilde{p}} - \omega_{\Tilde{p}'})t}\left(\sqrt{2}\abs{z_{\psi}}\right)^{\nu_{\psi}} e^{-\frac{i}{2}z_{\psi}^2}\left(\sqrt{2}\abs{z_{\phi}}\right)^{\nu_{\phi}} e^{-\frac{i}{2}z_{\phi}^2}.
\end{align}
\begin{align}
    \propto \int_{-\infty}^{t_0}dt\ e^{-ieEt^2+i(\omega_{\Tilde{p}} - \omega_{\Tilde{p}'})t}\left(-\sqrt{2eE}t\right)^{\nu_{\psi}+\nu_{\phi}}.
\end{align}        
The exponential factor in the integrand is a rapidly oscillating function, with the oscillation frequency growing towards larger values of $|t|$.  The main contribution to the integral is therefore given by the vicinity of the upper limit, $t\approx t_0$, so we can estimate
\begin{align}    
    I'\propto \int_{-\infty}^{t_0}dt\ e^{-ieEt^2+i(\omega_{\Tilde{p}} - \omega_{\Tilde{p}'})t}\left(-\sqrt{2eE}t\right)^{\nu_{\psi}+\nu_{\phi}}
    < \left(-\sqrt{2eE}t_0\right)^{\nu_{\psi}+\nu_{\phi}}\int_{-\infty}^{t_0}dt\ e^{-ieEt^2+i(\omega_{\Tilde{p}} - \omega_{\Tilde{p}'})t}
    \end{align}
    \begin{align}
    =\left(-\sqrt{2eE}t_0\right)^{\nu_{\psi}+\nu_{\phi}}\frac{\sqrt{\pi}}{2\sqrt{eE}}e^{i\frac{3}{4}\pi} e^{i\frac{(\omega_{\Tilde{p}} - \omega_{\Tilde{p}'})^2}{4eE}}\left(1-\text{erf}\left(e^{i\frac{\pi}{4}}\frac{(\omega_{\Tilde{p}} - \omega_{\Tilde{p}'})-2eEt_0}{2\sqrt{eE}}\right)\right).
\end{align}
By using Eq.\eqref{erf}, we have $I_{\text{upper}}'$, the upper limit of $I'$
\begin{align}
    I_{\text{upper}}'\propto& \left(-\sqrt{2eE}t_0\right)^{\nu_{\psi}+\nu_{\phi}}\frac{\sqrt{\pi}}{2eEt_0}e^{i\frac{3}{4}\pi} e^{-ieEt^2_0}
\end{align}
and
\begin{align}\label{4I'}
    \abs{I_{\text{upper}}'}\propto \abs{t_0}^{-2}.
\end{align}
We thus conclude that in the limit $t_0\to -\infty$ the amplitude (\ref{A2}) reduces to the transition amplitude in an eternal electric field.

\bibliographystyle{unsrt}
\bibliography{mybib}

\end{document}